\documentclass[a4paper,11pt]{article} 

\usepackage[latin1]{inputenc}
\usepackage[T1]{fontenc}
\usepackage{amsmath}
\usepackage{amssymb}
\usepackage{graphicx,subfig}

\newtheorem{teorema}{Theorem}
\newtheorem{proposizione}{Proposition}
\newtheorem{corollario}{Corollary}
\def\min{{\rm min}}
 \def\eps{ { \varepsilon } }
\let\flusso=\phi       
\def\phi{{\varphi}}

\newcommand{\dif}{\mathrm{d}}
\def\naturali{{\bf N}}
\def\Kscr{{\cal K}}
\def\gt{>}
\def\lt{<}
\newcommand{\equal}{\buildrel {\rm def} \over {=} }
\newcommand{\cov}{\mathrm{cov}}
\newcommand{\corr}{\mathrm{r}}
 
\newcommand{\blank}{\vskip 1ex \noindent}

\title{A series expansion for the time autocorrelation 
  of dynamical variables }

\author{A.M.~Maiocchi \and A.~Carati \and	A.~Giorgilli}

\date{\today}

\begin{document}

\maketitle

\begin{abstract}
We present here a general iterative formula which gives a (formal)
series expansion for the time autocorrelation of smooth dynamical variables,
for all Hamiltonian systems endowed with an invariant measure. We add
some criteria, theoretical in nature, which enable one 
to decide whether the decay of the correlations is exponentially fast
or not. One of these criteria is implemented numerically for the case of the
Fermi-Pasta-Ulam system, and we find indications which might suggest
a sub-exponential decay of the time autocorrelation of a relevant
dynamical variable. 
\end{abstract}

\section{Introduction}
It is well known that one of the most important indicators of the
chaotic behaviour of a dynamical system is the rate of decay to
zero of the time autocorrelation of dynamical variables. A vast
literature exists on this subject, mainly addressed to systems
of some special class (for example, Anosov systems, see
\cite{liverani}, or systems with few degrees of freedom, such as
billiards, see \cite{chernov}). In this paper we study the decay of
correlations in the general frame of the dynamical theory of Hamiltonian
systems, with an invariant measure. Such an approach has already
provided interesting results, as it allows to obtain results valid
in the thermodynamic limit (see
\cite{carati,maiocchicarati,caratimaiocchi}).

Here we provide a power series expansion (with respect to time) of the
time autocorrelation of a smooth function $f$
(Theorem~\ref{teor:serie}, Section~\ref{sez:generale}). It turns out
that the coefficients of such a series are essentially the variances of
$L_H^nf$, where $L_H$ is the Poisson bracket operator relative to
the Hamiltonian $H$. We then establish a necessary and sufficient
condition for its time-decay not to be exponential (Corollary~\ref{cor:analitico},
Section~\ref{sez:serie}). We finally give a numerical application to
the case of the Fermi-Pasta-Ulam system (FPU). By truncating the series
up to order 12, for systems up to 364 degrees of freedom, we
approximate the poles of the Laplace transform of the time
autocorrelation of a suitably chosen dynamical variable. Some comments
are made on the possibility that such approximations give indications for a
sub-exponential decay of the autocorrelation.

In Section~\ref{sez:generale} the relevant notions on time correlation
functions are recalled. In the same section we give a theorem which
establishes a kind of continuity (with respect to a suitable distance) of time
correlations in the $L^2$ space of the dynamical
variables, and also Theorem~\ref{teor:serie} for the series expansion.

Section~\ref{sez:serie} is devoted to a study of some general
properties of the series expansion. Here an interesting link
with the Stieltjes moment problem is pointed out (see
Theorem~\ref{teor:stieltjes}), and a necessary and sufficient
criterion for the  time-decay of the autocorrelation to be
exponential is given as Corollary~\ref{cor:analitico}. A closer
examination of this problem is given in Appendix~\ref{app:hausdorff}.

Then, the
results of some numerical computations on the time autocorrelations of
a significant variable in a FPU chain are reported in
Section~\ref{sez:numerica}. Finally, in Section~\ref{sez:conclusione}
some comments on the usefulness of the present method are also given.
Two more Appendices complete the paper.

\section{Time correlations in their ``natural'' space}\label{sez:generale}
We recall here some standard concepts within the measure theoretic approach to
dynamical systems.
We consider a Hamiltonian system on a phase space $\mathcal{M}$
endowed with a probability measure $\mu$ invariant with
respect to the time flow $\flusso^t$ induced by the Hamiltonian $H$.
We also consider the time evolution operator $\hat{U}_t$ acting on
the space of the square integrable functions from $\mathcal{M}$ to
$\mathbb{R}$, i.e., $L^2(\mu, \mathcal{M})$.  The operator
$\hat{U}_t$ maps $f$ to $\hat{U}_t f = f\circ \flusso^{-t}$. It is known
after Koopman (see \cite{koopman}) that $\hat U_t$ defines a
one-parameter group of unitary operators, namely operators preserving
the norm $\|\cdot \|$ in $L^2(\mu, \mathcal{M})$.

We are interested in the time autocorrelation of a dynamical
variable $f\in L^2(\mu, \mathcal{M})$, which is defined as
\begin{equation}\label{eq:def_correlazione}
\mathbf C_f(t)\equal \langle f_t\,f\rangle - \langle f \rangle^2=\left\langle
\left( f_t-\langle f\rangle\right)\left( f-\langle f\rangle
\right)\right\rangle\ ,
\end{equation}
where $f_t$ is a shorthand for $\hat U_t f$, and $\langle \cdot\rangle$
denotes mean value with respect to the probability measure $\mu$.

It will also be useful to consider the time correlation between two
dynamical variables $f$ and $g$, defined as
$$
\mathbf C_{f,g}(t) \equal  \left \langle f_t g \right\rangle- \langle f
\rangle \langle g\rangle =\left\langle \left(f_t-\langle f\rangle\right)
\left(g-\langle g\rangle\right)\right\rangle\ .
$$

It is well known from probability theory that the concepts of variance
and correlation acquire a geometrical meaning if the covariance
between two random variables $f$ and $g$ is used as a scalar product.
The covariance is defined as
$$
\cov(f,g) \equal \left\langle \left(f-\langle f\rangle\right)
\left(g-\langle g\rangle\right)\right\rangle= \mathbf C_{f,g}(0)\ ,
$$
This choice leads to take as norm\footnote{We can adopt this
  quantity as a norm in strict sense only if all dynamical variables
  which differ by a constant are identified.  For a function with
  zero mean the covariance $\cov(f,f)$ actually coincides with the
  usual $L^2$ norm.  In this paper we shall use both norms, keeping
  the usual symbol $\|\cdot\|$ for the  $L^2$ norm.} of a dynamical
variable $f$ its standard deviation $\sigma_f$, defined by
$$
\sigma^2_f\equal \cov(f,f)= \left\|f\right\|^2 - \left\langle
f\right\rangle^2=  \left\|f-\langle f\rangle \right\|^2\ ,
$$ 
$\sigma^2_f$ being the variance of $f$.

The notion that some relation exists between two random
variables $f$ and $g$ is made quantitative through the correlation
coefficient $\corr(f,g)$ defined as 
$$
\corr(f,g)\equal \cov(f,g)/\sigma_f \sigma_g\ .
$$
Using the Schwarz inequality it is easily seen in our case that one has
$|\cov(f,g)|\le \sigma_f \sigma_g$,   so that
$|\corr(f,g)|\le 1$.
Two variables are orthogonal in this metric if
they are uncorrelated, i.e., if $\corr(f,g)=0$, and collinear if
$\corr(f,g)=\pm 1$.  

We also emphasise that, in view of the above definitions, the time
autocorrelation of a function $f$ is nothing but the covariance between $\hat
U_t f$ and $f$. Therefore one immediately has
\begin{equation}\label{eq:correlazione_e_varianza}
\mathbf C_f(0)=\sigma^2_f\quad \mbox{and }
\left|\mathbf C_f(t)\right| \le \sigma_f^2\ .
\end{equation}
Thus, the variance of a dynamical variable is the natural scale of its
time autocorrelation.

\blank We come now to state two useful theorems.  The first one points
out an interesting property of the time correlation.  The mere fact
that two dynamical variables are strongly correlated (i.e.,
sufficiently collinear in the scalar product given by the covariance)
entails a similar behaviour of their time autocorrelations.
Equivalently, we can say that in the neighbourhood of each dynamical
variable $f$, i.e., in the set of all dynamical variables strongly
correlated with $f$, the behaviour of the time autocorrelations is
determined by that of $\mathbf C_f(t)$.

\begin{teorema}\label{teor:correlate}
Let  $\mu$ be a probability measure on the phase space
$\mathcal{M}$, invariant for the flow generated by $H$, and let $f$
and $g$ be dynamical variables belonging to $L^2(\mu, \mathcal{M})$
such that one has $|\corr(f,g)|\ge
1-\eps^2/2$ for some $\eps>0$. Then there exists a multiple $\tilde
f=\alpha f$ of $f$, with $\alpha\equal (\mathrm{sign} (\corr(f,g))
\sigma_g/\sigma_f)$, such that 
\begin{equation}\label{eq:maggiorazione_correlazione}
\bigl|\mathbf C_{\tilde f,g}(t)-\mathbf C_{\tilde f}(t)\bigr|\le
\eps \sigma_g^2
\end{equation}
and
\begin{equation}\label{eq:maggiorazione_autocorrelazione}
\bigl|\mathbf C_g(t)-\mathbf C_{\tilde f}(t)\bigr| \le \left(\eps^2+2 \eps\right)
\sigma^2_g\ .
\end{equation}
\end{teorema}

\noindent
\textbf{Proof.} Both inequalities come from the remark that
$$
\sigma_{g -\tilde f}= \sigma^2_g+\sigma^2_{\tilde f}- 2 \cov(\tilde
f,g) = 2\sigma^2_g-2 \sigma^2_g \corr(f,g) \le \eps^2 \sigma^2_g\ ,
$$
which is due to the identities $\sigma_{\tilde f}=\sigma_g$ and
$\corr( \tilde f,g)= \bigl|\corr(f,g)\bigr|$. In fact,
(\ref{eq:maggiorazione_correlazione}) hence follows by noting that
$$
\bigl|\mathbf C_{\tilde f,g}(t)-\mathbf C_{\tilde f}(t)\bigr|= \bigl|\cov(\tilde f_t,g)-\cov(\tilde
  f_t,\tilde f)\bigr|=\bigl|\cov(\tilde f_t,g-\tilde f)\bigr|\le \sigma_{g\phantom{\tilde f}}\!\!
\sigma_{g-\tilde f}\ ,
$$
while, in a similar way, (\ref{eq:maggiorazione_autocorrelazione})
comes from the following relations
\begin{equation*}
\begin{split}
\bigl| \mathbf C_g(t)-\mathbf C_{\tilde f}(t)\bigr| &=
\bigl|\mathbf C_{g-\tilde f}(t)+\mathbf C_{g,\tilde f}(t)+\mathbf C_{\tilde
f,g}(t)-2\mathbf C_{\tilde f}(t) \bigr|\\
&\le \sigma^2_{g-\tilde f}+2
\sigma_{g\phantom{\tilde f}}\!\!\sigma_{g-\tilde f}\ .
\end{split}
\end{equation*}
Here, in the second line, use is made of the previous inequality and
of the identity $\mathbf
C_{g,\tilde f}(t)=\mathbf C_{\tilde f,g}(-t)$, due to the invariance
of the measure.
\begin{flushright}Q.E.D.\end{flushright}
\blank

A central role in this paper will be played by the formal power series
expansion of $\mathbf C_f(t)$ with respect to time, which is given in
the following theorem. Use will be made of the definition of the
$k$-th order Lie derivative of $f$, for $k\ge 1$, namely
$f^{(k)}\equal[f^{(k-1)},H]$, where $f^{(0)}\equal f$ and
$[\cdot,\cdot]$ denotes the Poisson brackets.

\begin{teorema}\label{teor:serie}
Let $\mu$ be a probability measure on the phase space 
$\mathcal{M}$, invariant for the flow generated by $H$, and let
$n>0$. Then, for any dynamical variable $f\in L^2(\mu, \mathcal{M})$
such that $f^{(k)}\in  L^2(\mu, \mathcal{M})$ for all $k\le n$, one has
\begin{equation}\label{eq:serie_troncata}
\begin{split}
\mathbf C_f(t) = &\sigma^2_f
 +\sum_{k=1}^{n}(-)^k\bigl\|f^{(k)}\bigr\|^2\frac{t^{2k}}{(2k)!}\\ &+
 (-1)^{n+1} \int_0^t\dif t_1 \ldots
 \int_0^{t_{2n-1}} \dif t_{2n} \bigl \|f_{t_{2n}}^{(n)}-f^{(n)}\bigr\|^2\ . 
\end{split}
\end{equation}
\end{teorema}

\noindent
\textbf{Proof.}
We use the simple chain of identities
\begin{equation}\label{eq:correlazione_inversa}
  \|f_t-f\|^2=2\sigma^2_f-2\mathbf C_f(t)= 2 \|f\|^2
  -2 \langle f_t\,f\rangle\ ,
\end{equation}
which hold true for any invariant measure, and the remark that
\begin{equation}\label{eq:evoluzione_differenza}
f_t-f=\int_0^t \hat{U}_{t-s}\dot{f}\dif s,
\end{equation}
where $\dot{f}=[f,H]$. One can check equations
(\ref{eq:correlazione_inversa}) by writing down the square at the
l.h.s., while (\ref{eq:evoluzione_differenza}) comes 
\footnote{See \cite{carati} for a more detailed proof.} from the
variation of constants formula applied to $f_t-f$.

We go on by writing the l.h.s. of
(\ref{eq:correlazione_inversa}) as
\begin{eqnarray}
\|f_t-f\|^2 &=& \int \dif\mu \int_0^t\dif t'\int_0^t\dif t''
\bigl(\hat{U}_{t-t'}\dot{f}\bigr)\bigl(\hat{U}_{t-t''}\dot{f}\bigr)\nonumber\\
&=&\int \dif\mu \int_0^t\dif t'\int_0^t\dif
t''\bigl(\hat{U}_{t'-t''}\dot{f}\bigr)\dot{f}\label{passo_1}\\
&=&2\int_0^t \dif t'\int_0^{t'} \dif t''\left(\bigl\|\dot{f}\bigr\|^2 -
\frac{1}{2} \bigl\|\dot{f}_{t''}-\dot{f}\bigr\|^2\right),\label{passo_2}
\end{eqnarray}
where, in the second line, the invariance of the measure with respect
to the time evolution is used, so that the equality proceeds from a
simple change of coordinates, and in the last line the same argument
is used, as well as relation (\ref{eq:correlazione_inversa}) applied
to the variable $\dot f$. The order of the integrals can be exchanged
if $\|\dot{f}\|$ is finite, because the integral (\ref{passo_1}) is
absolutely convergent in this case, in virtue of Schwarz inequality.
In~(\ref{passo_2}) we repeat the same calculation for
$\|\dot{f}_{t''}-\dot{f}\|^2$, using $\dot f = [f,H]$. This gives us
the relation
\begin{eqnarray*}
\left\|f_t-f\right\|^2 &= &2\left(\bigl\|\dot{f}\bigr\|^2\frac{t^2}{2!}-\bigl\|f^{(2)}\bigr\|^2\frac{t^4}{4!}\right)\\
& &+\int_0^t\dif t_1\int_0^{t_1}\dif t_2\int_0^{t_2}\dif
t_3\int_0^{t_3}\dif t_4\bigl\|f^{(2)}_{t_4}-f^{(2)}\bigr\|^2\ ,
\end{eqnarray*}
where $f^{(2)} = [\dot f,H]$.  The same procedure can be iterated up
to $k=n$ by
recursively defining
$f^{(k)}\equal [f^{(k-1)},H]$, starting with $f^{(0)}\equal f$, and
using the hypothesis that
$f^{(k)}\in L^2(\mu, \mathcal{M})$.  This completes the proof.
\begin{flushright}Q.E.D.\end{flushright}
\blank

\noindent
\textbf{Remark 1.} Theorem~\ref{teor:serie} suggests that,
at least in a formal way, one can express the time autocorrelation of
a dynamical variable $f$ as a power series with respect to time
\begin{equation}\label{eq:serie}
  \mathbf C_f(t)=\sigma^2_f
  +\sum_{k=1}^{+\infty}(-)^k\bigl\|f^{(k)}\bigr\|^2\frac{t^{2k}}{(2k)!}\ .
\end{equation}
To simplify the notations, we define
\begin{equation}\label{eq:definizioni_a_e_c}
c_0\equal \sigma_f^2 \ ,\quad c_n\equal \bigl\|f^{(n)}\bigr\|^2 \mbox{
  for } n> 0\ .
\end{equation}
We point out that all these coefficients are positive. 
We could have got to formula (\ref{eq:serie}) also by expressing $f_t$
as a time power 
series through its Lie derivatives and by integrating $\langle
f_t\,f\rangle$  by parts, taking into account the observation
that the mean value of any function which can be written as $[g,H]$
vanishes\footnote{For the same reason, $\sigma_{f^{(k)}}$ can replace
  $\|f^{(k)}\|$ in all the previous formulae. We will keep the given
  notation, because this way it is more straightforward to understand
  how to compute the coefficients of the series.}, since $\bigl\langle
[g,H]\bigr\rangle=\frac{\dif}{\dif t} \langle g\rangle$.
\blank

\noindent
\textbf{Remark 2.} We emphasise that the term in the
second line in equation (\ref{eq:serie_troncata}), i.e., the
remainder, turns out to be positive or negative according to
$n$ being odd or even, respectively. Therefore, the truncations of
series (\ref{eq:serie}) provide upper bounds for the time
autocorrelation of $f$ if truncated at an odd $n$, and lower bounds if
truncated at an even $n$. Such bounds provide some information on the
behaviour of the time autocorrelation of $f$ for finite times.
The simplest example is that of the first order truncation, which
proved to be very helpful for the study of relaxation times in a
Hamiltonian setting (see \cite{carati,maiocchicarati,caratimaiocchi}).
\blank

As a final remark, we point out that some \textit{a priori}
restrictions on the norms of the successive derivatives come from the Schwarz
inequality~(\ref{eq:correlazione_e_varianza}). Indeed, one has the
following Proposition, whose proof is deferred to Appendix~\ref{app:a_priori}.
\begin{proposizione}\label{prop:a_priori}
The coefficients $c_n$ defined by
(\ref{eq:definizioni_a_e_c}) are such that the polynomials $P_n^l(y)$,
$Q_n^l(y)$ defined as
\begin{equation}\label{eq:polinomi_generalizzati}
P_n^l(y) \equal  2 a_0^l+\sum_{k=1}^{2n}(-)^ka_k^l y^k\ ,\quad
Q_n^l(y) \equal  \sum_{k=0}^{2n}(-)^ka_{k+1}^l y^k\ ,
\end{equation}
with
\begin{equation}\label{eq:coefficienti_generalizzati}
a_n^l\equal c_{n+l}/(2n)!\ ,
\end{equation}
are positive semi-definite, for any $n,l\ge 0$.
\end{proposizione}

\noindent
The statement of Proposition~\ref{prop:a_priori}
entails some restrictions on the coefficients $c_n$, since $P_n^l(y)$ and
$Q_n^l(y)$ are two-indexed sequences of positive semi-definite
polynomials.  Unfortunately, such relations can be expressed only in a
quite complicated way, through the Jacobi-Borchardt theorem
(see, for example, \cite{obre}), which gives some relations among the
roots of the polynomials, and Newton's identities on the relations
between the coefficients of a polynomial and its roots.

\section{Asymptotic behaviour of the correlations}\label{sez:serie}
In many cases one is interested in
determining whether the time correlation tends to zero as $t\to \infty$
and, if this is the case, which is the law controlling the decay. Indeed, it
is commonly stated that if the correlations decay exponentially fast
(for a suitable wide class of functions)
then the system is chaotic.
The decay is linked to the analyticity property of the Fourier
transform of the correlations (see for example \cite{ruelle}): in
particular it is  well known that the Fourier transform is analytic in
a strip of width $1/\tau$ if and only if the correlation decays at least as
$\exp(-t/\tau)$. We recall that the correlations can always be
expressed as the Fourier transform of an even positive measure, the so
called ``spectral measure'', namely as
\begin{equation}\label{eq:bochner}
\mathbf C_f(t) = \frac 1{2\pi}\int_{-\infty}^{+\infty} e^{i\omega
  t} \dif \alpha(\omega) \ ,
\end{equation}
where $\alpha$ is a positive Borel measure on $\mathbb{R}^+$ (see,
for example, \cite{feller}).


On the other hand, it is not so easy to get information about
the spectrum: we devise to get it using the Laplace transform $F(s)$
of the time autocorrelation, namely,
\begin{equation}\label{eq:trasformata_funzione}
F(s) \equal \int_0^{+\infty} e^{-st}\mathbf C_f(t) \quad \mbox{for }
s\in \mathbb C\ .
\end{equation}
It is well known that the Laplace transform of a
function which decays faster than $\exp(-t/\tau)$ is analytic in the
half-plane $\operatorname{Re} s>-1/\tau$, so that, again, the control
on the decay of correlations in linked to the analyticity properties
of $F(s)$, as in the case of the Fourier transform.
The great advantage of the latter approach lays in the fact that the
Laplace transform of the correlation is essentially the Stieltjes
transform of the spectral measure. This is stated as
Theorem~\ref{teor:stieltjes} below.

The point is that, as first shown by Stieltjes (see
\cite{stieltjes}), if one knows the asymptotic expansion
in powers of $1/s$ of a functions $F(s)$ which is the Stieltjes
transform of a positive measure, then one can recover $F(s)$ itself
through a convergent scheme of continued fractions expansions, defined
in terms of the coefficients of the asymptotic expansion. In other
words, one can construct a rational approximation
$$
F_n(s)=\frac{P_n(s)}{Q_n(s)}\ ,
$$
which converges uniformly in $\operatorname{Re} s>0$, as $n\to
+\infty$, to $F(s)$. For
the case of the time autocorrelation $\mathbf C_f(t)$, the asymptotic
expansion is obtained just by integrating the expansion
\eqref{eq:serie} term by term. In fact, the 
Laplace transform \eqref{eq:trasformata_funzione} can be formally written as 
\begin{equation}
F(s)= \int_0^{+\infty}\left(\sum_{n=0}^{+\infty}(-)^n\frac{c_n}{(2n)!}t^{2n}
\right)e^{-st}\, \dif t =\sum_{n=0}^{+\infty}
(-)^n\frac{c_n}{s^{2n+1}}\ . \label{eq:trasformata_serie}
\end{equation}
This expansion would give a convergent expansion,
provided one can exchange the sum of the series with the
integration over time. This is in general not permissible, in view
of the fact that the expansion in power series of time converges (in
general) only in a circle of finite radius and not up to infinity. 
However,  it is easy to show that series (\ref{eq:trasformata_serie})
nevertheless provides the asymptotic expansion of $F(s)$ we are
looking for. One has to use formula \eqref{eq:serie_troncata} and
check that the remainder grows at most as a power of time.

We are now able to state the main result of this section, namely,
the following theorem, together with a relevant corollary on the
decay of the correlations.
\begin{teorema}\label{teor:stieltjes}
Let $\mathbf C_f(t)$ be analytic in $t$ about the origin and
continuous for any $t\in \mathbb R$. Then the following statements hold:
\begin{enumerate}\label{prop:stieltjes_1}
\item in the half-plane $\operatorname{Re} s> 0$ the Laplace transform
  $F(s)$ of $\mathbf C_f(t)$ is analytic and one has
\begin{equation}\label{eq:stieltjes_laplace}
F(s) = \frac {s}{\pi} \int_0^{+\infty} \frac {\dif \alpha(\omega)}{s^2+\omega^2} \ ,
\end{equation}
where $\alpha(\omega)$ is the positive Borel measure such that
\eqref{eq:bochner} holds;
\item\label{prop:stieltjes_2} the continued fraction which
  approximates the asymptotic series (\ref{eq:trasformata_serie})
converges to $F(s)$ for $\operatorname{Re} s> 0$ and the Borel
positive measure $\Phi(u)$ defined on $\mathbb R^+$ by $\Phi(u)\equal(1/\pi)
\alpha(\sqrt u)$ solves the Stieltjes moment problem for the coefficients
$c_n$ defined by (\ref{eq:definizioni_a_e_c}). Moreover,
$\alpha(\omega)/\pi$ solves the symmetric Hamburger moment problem with
moments $c'_{2n}=c_n$, $c'_{2n+1}=0$.
\end{enumerate}
\end{teorema}

\begin{corollario}\label{cor:analitico}
$\mathbf C_f(t)$ decays exponentially fast as $t$ goes to infinity if
  and only if the coefficients $c_n$ are such that $\alpha'(\omega)$
  exists and $\sqrt{u}\Phi'(u)$ is analytic. 
\end{corollario}

\noindent
\textbf{Remark 3.}We will state our results in
  terms of the regularity properties of $\alpha$ or $\Phi$ according
  to convenience. We prefer the study of the properties of
  $\Phi$, when the convergence properties of Stieltjes continued
  fraction approximation are needed.
\blank
\noindent
\textbf{Proof of Theorem~\ref{teor:stieltjes}.}
First of all, we observe that (\ref{eq:stieltjes_laplace}) follows by
expressing $\mathbf C_f(t)$ via (\ref{eq:bochner}) in the definition
(\ref{eq:trasformata_funzione}) of $F(s)$ and exchanging the order of
integration. This can be done for any $s$ for which $\operatorname{Re}
s>0$, on account of the Tonelli-Fubini theorem for the Stieltjes integral
(see \cite{widder}), and it is well known that the so defined $F(s)$
is analytic in the same region.

We come now to the proof of statement~\ref{prop:stieltjes_2}.
We notice that the form of series (\ref{eq:trasformata_serie}) recalls
the Stieltjes work \cite{stieltjes} on the link between the asymptotic
series, their expansion in continued fractions and the Stieltjes transform. 
In dealing with this subject he met with the
moment problem that now bears his name. We recall that, given
a sequence of positive numbers $c_n$, the Stieltjes moment problem
consists in checking whether there exists a measure $\Phi$ on
$\mathbb{R}^+$ such that
\begin{equation}\label{eq:integrale_momenti_stieltjes}
\int_0^{+\infty}u^n\dif \Phi(u)=c_n\ .
\end{equation}
The solution exists if and only if
\begin{equation}\label{eq:soluzione_stieltjes}
\det (\Delta_n)\ge 0 \quad \mbox{and} \quad \det(\tilde{\Delta}_n)\ge 0\quad
\forall n\ ,
\end{equation}
where the matrices $\Delta_n$, $\tilde{\Delta}_n$ are defined by
\begin{equation}\label{eq:def_determinanti}
\Delta_n\!\equal\!\left[\!
\begin{array}{cccc}
c_0 & c_1 & \cdots &c_n\\
c_1 & c_2 &\cdots & c_{n+1}\\
\vdots& \vdots & \ddots& \vdots\\
c_n & c_{n+1} &\cdots & c_{2n}
\end{array}
\!\right] \ ,\ 
\tilde{\Delta}_n\!\equal\! \left[\!
\begin{array}{cccc}
c_1 & c_2 & \cdots &c_{n+1}\\
c_2 & c_3 &\cdots & c_{n+2}\\
\vdots& \vdots & \ddots& \vdots\\
c_{n+1} & c_{n+2} &\cdots & c_{2n+1}
\end{array}
\!\right]\ ,
\end{equation}
and if the equality in (\ref{eq:soluzione_stieltjes}) is obtained for
$\Delta_k$, then it is obtained for $\Delta_n$, with $n\ge k$ and for
$\tilde{\Delta}_n$, with $n\ge k-1$.
Furthermore, the solution is unique if there exists $D>0$ such that
\begin{equation}\label{eq:convergenza}
  c_n\le D^n(2n)!\ ,
\end{equation}
holds (for a proof with more recent methods, see \cite{ahiezer}).
We point out that, in our case, since $C_f(t)$ is analytic about the
origin,  there should exist $D>0$ such that the previous condition is
satisfied, in view of the Cauchy-Hadamard criterion.

Coming back to our problem, we collect the results of paper
\cite{stieltjes} of interest to us in the following
statement: \textit{if the function $F(s)$ admits the representation
  (\ref{eq:stieltjes_laplace}), then the continued fraction which
  approximates the asymptotic expansion (\ref{eq:trasformata_serie})
  converges to $F(s)$ for $\operatorname{Re} s>0$; moreover, the
  Stieltjes moment problem for the sequence $c_n$
  is soluble.}

The previous statement is proved this way. In equation
(\ref{eq:stieltjes_laplace}) we put  $z\equal s^2$, $u\equal
\omega^2$, $\Phi(u)\equal\alpha(\sqrt u)$. Then we have the formal
chain of equalities
\begin{equation}\label{eq:sviluppo_serie}
s\int_0^{+\infty} \frac{\dif
  \Phi(u)}{z+u}=F(s)=s\left(\frac{c_0}{z}-\frac{c_1}{z^2}+\frac{c_2}{z^3}
+\ldots\right)
\ , 
\end{equation}
where again $\Phi$ is a Borel positive measure on $\mathbb R^+$. Stieltjes
showed that the
continued fraction which approximates the term in brackets at the
r.h.s. converges to the integral at the l.h.s, and that the
Stieltjes moment problem for the sequence $c_n$ has a solution for the
measure $\Phi$. Then it is straightforward to check that $\alpha/\pi$
solves the associated symmetric Hamburger moment problem, which means
that
$$
\frac 1\pi \int_{-\infty}^{+\infty} \omega^{2n}\dif \alpha(\omega)=
c_n\ ,\quad \frac 1\pi \int_{-\infty}^{+\infty} \omega^{2n+1}\dif
\alpha(\omega)=0\ .
$$
\begin{flushright}Q.E.D.\end{flushright}
\blank

\noindent
\textbf{Remark 4.} Notice that we have incidentally shown that the
  coefficients $c_n$ are such that conditions
  (\ref{eq:soluzione_stieltjes}) are satisfied at any order. This
  restriction on the values of the coefficients,
  together with  the analogous ones imposed by the application of the
  previous theorem to the dynamical variables $f^{(n)}$, for all $n$,
  already contains the restriction given by
  Proposition~\ref{prop:a_priori}. However, we point out that
  Proposition~\ref{prop:a_priori} holds up to a given order even if
  the existence of $f^{(n)}$ beyond such an order is not guaranteed.
\blank

In view of Corollary~\ref{cor:analitico}, the question of the
asymptotic behaviour of the time autocorrelations can be restated as a
regularity problem for the measure  $\Phi$, which solves Stieltjes
moment problem for the $c_n$, or, equivalently, for $\alpha$ which
solves the associated Hamburger moment problem. An answer to this
question can be given only if the whole sequence of moments 
$c_n$ is known. However, it is also of interest to understand which is the
qualitative difference between the approximation (at a given order)
of an analytic measure and 
that of a measure with at least a singular point. We try here to
give an answer to the latter question, deferring to
Appendix~\ref{app:hausdorff} the discussion of
which is the link between the behaviour of the moments and the
regularity of the solution (see Proposition~\ref{prop:ahiezer},
\ref{prop:hausdorff}, \ref{prop:convergente} therein).

We have already pointed out that a rational approximation $F_{2n}(s)$ of the
Laplace transform $F(s)$ can be constructed from
(\ref{eq:sviluppo_serie}) by standard methods (see
Chapter~2 of Stieltjes memoir \cite{stieltjes}), which give
$$
F_{2n}(s)=s\frac{P_{2n}(s)}{Q_{2n}(s)}=s\sum_{k=0}^n \frac{\rho_k}{s^2+
\omega^2_k}\ .
$$
Such a function, in turn, can be rewritten as the following Stieltjes integral
$$
F_{2n}(s)=s \int_0^{+\infty} \frac{\dif \alpha_n(u)}{s^2+u}\ ,
$$
in which the measure $\alpha_n$ is piecewise constant and has a jump
$\rho_k$ at the points $u=\omega^2_k$.
Therefore, in order that the limiting measure $\alpha=\lim \alpha_n$ be
continuous, it is necessary that, as the approximation order grows,
the poles become dense and their residues tend to zero. As a
consequence, \emph{we will take as a qualitative indication of a
  sub-exponential decay the property that the rational approximations
  have at least a pole which remains isolated, with a stable residue,
  as the order grows}.

\section{Numerical study of the FPU chain}\label{sez:numerica}
Checking whether the spectral measure is analytic, or even whether the
criterion of Proposition~\ref{prop:ahiezer} of
Appendix~\ref{app:hausdorff} is satisfied for a concrete problem is a 
major task, because it requires information on an infinite number of
coefficients.  Thus, as a first attempt to implement the ideas of
Section~\ref{sez:serie}, we computed numerically some
coefficients $c_n$ for a model which is widely studied in the
literature and is of great relevance for the foundations of
statistical mechanics, namely the Fermi-Pasta-Ulam model (FPU, see
\cite{fpu}).  It describes a one-dimensional chain of $N+1$ particles
interacting through nonlinear springs. The Hamiltonian of such a
system, after a suitable rescaling, can be written as
$$
H=\sum_{j=0}^N \frac{p_j^2}{2} +\sum_{j=0}^{N-1}\left[
  \frac{\left(q_j-q_{j+1}\right)^2}{2}+
  \frac{\alpha\left(q_j -q_{j+1}\right)^3}{3} +
\frac{\beta\left(q_j-   q_{j+1}\right)^4}{4}\right]\ ,
$$ 
where $p=(p_0,\ldots ,p_N)$ and $q=(q_0,\ldots ,q_N)$ are
canonically conjugated variables, while $\alpha$ and $\beta$ are
parameters that control the size of the nonlinearity. We choose as
invariant probability measure the Gibbs one, i.e., $\dif \mu
(p,q)\equal Z^{-1}(T) \exp(-H(p,q)/T)\dif p\,\dif q$, where $Z(T)$ is
the partition function and $T>0$ the temperature, and for the purpose
of speeding up the numerical evaluation of the integrals involving the
measure, we consider here a chain with one end point fixed and the
other one free, i.e, we impose only the boundary condition $q_0=0$, $p_0=0$ (see
footnote~\ref{nota:indipendenti} below).  We note however that a few
computations made on the model in which
both end-points are fixed have shown no significant difference.  
Concerning the parameters, we consider the
often studied case in which $\alpha=\beta=1/4$.  The relevant fact is
that  as $T$ approaches 0 the contribution due to the
nonlinear terms becomes statistically negligible.

It is well known that if the nonlinear terms are neglected then the
Hamiltonian is integrable, admitting $N$ 
normal modes of oscillation, and that the motion is quasi periodic so
that, for every 
$k=0,\ldots,N-1$, the energy $E_k$ of the $k$-th mode is a constant
of motion (and thus, as obvious, the time autocorrelation of such a
dynamical variable is constant).
Our aim is to investigate what happens when the nonlinearity is
introduced as a perturbation, namely, for $T$ positive but close to 0.
We are particularly interested in the exchange of energy among the
normal modes.  To this end, a good choice could be to consider the
fraction of energy localized on the low frequency modes, defined, for
example, as
$$
\mathcal{E}=\frac 1N\sum_{k=0}^{N/2}E_k\ ,
$$
since its time variation is a convenient indicator of the flow of
energy from low to high frequency modes, and vice versa.
As a numerical tool we want to compute numerically the coefficients of the series
(\ref{eq:serie}) for this function, since, in view of
Theorem~\ref{teor:stieltjes}, they give indications on the asymptotic
behaviour of the autocorrelation.

As a matter of fact, we notice that the function $\mathcal{E}$ may not
represent the best choice.  For,  such a function 
is strongly correlated with the Hamiltonian,
so that, in virtue of Theorem~\ref{teor:correlate}, its
autocorrelation remains close to that of the Hamiltonian, thus
remaining far from zero.  So, we rather took a
suitable modification of $\mathcal{E}$, namely,
the projection of $\mathcal{E}$, via the Gram-Schmidt orthogonalization
process, on the space of the dynamical variables
uncorrelated with $H$.   Thus we consider the dynamical variable
$\tilde{\mathcal{E}}\equal \mathcal{E}- \cov(\mathcal{E},H) 
H/\sigma^2_H$. 

Our calculation proceeds as follows.
We extract a sample in phase space according to the Gibbs
measure\footnote{We extract $p_j$ and $r_j\equal q_j-q_{j-1}$, for
  $j=1,\ldots,N$, which are stochastically independent variables for
  the Gibbs measure with respect to the present Hamiltonian. Here lies
  the great advantage of studying the chain with one end-point
  free.\label{nota:indipendenti}}
and we estimate  the $L^2$-norm $\|\tilde{\mathcal{E}}^{(n)}\|^2$ of the functions
$\tilde{\mathcal{E}}^{(n)}=[\tilde{ \mathcal{E}}^{(n-1)},H]$ generated by $\tilde{
\mathcal{E}}^{(0)}=\tilde{ \mathcal{E}}$ taking the mean value on our
sample. For this purpose, we observe that $\tilde{\mathcal{E}}^{(n)}$
is a certain function of $p$ and $q$, which can be expressed via Fa\`a di
Bruno's formula for the derivatives of any order of a composed
function. This formula requires the computation of some combinatorial
coefficients, which needs a smart implementation in order to reduce the
computation time. We report in Appendix~\ref{app:faa} the scheme we
have followed.

In order to study the Laplace transform of the time autocorrelation of
$\tilde{ \mathcal{E}}$, following the procedure of Stieltjes, we approximate it up
to order $n$ with  rational functions, and look for the poles of such an
approximating function in the complex plane (see chapter 2 of the memoir
\cite{stieltjes}). Our aim is to compare the behaviour of the time 
autocorrelation of $\tilde{ \mathcal{E}}$ with some functions which behave in a
somehow known way, by comparing the poles of the corresponding
rational approximations of the Laplace transforms. The first
comparison function is the time autocorrelation of a variable which should
lose correlation in a very short time, i.e., 
(the orthogonal projection of) the kinetic energy $\tilde K$ of the
first  half of the chain's particles. The second one will be a given
function of time which has an exponential decay (see below for its
actual form).
\begin{figure}[th]
\centering
\includegraphics[width=0.75\textwidth]{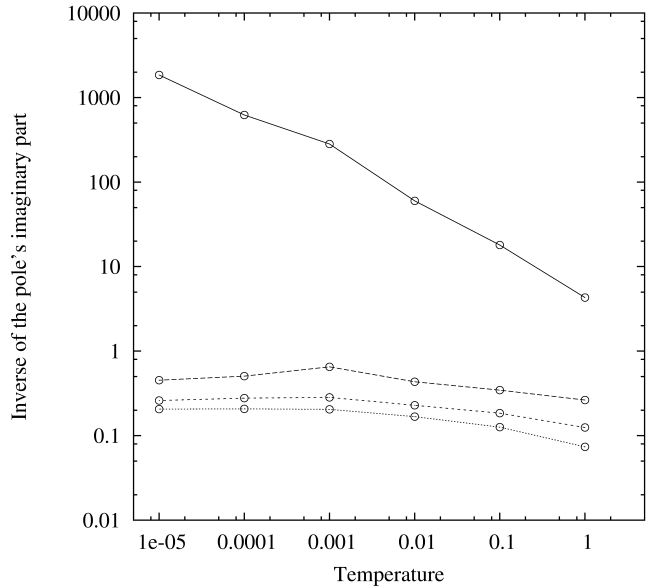}
\caption{\label{fig1} Inverse of the imaginary part of the four
    poles of the Laplace transform of the time autocorrelation of $\tilde{
    \mathcal{E}}$ versus temperature, in logarithmic scale. Here
    the number of particles is 
    $N=40$.}
\end{figure}
\begin{figure}[th]
\centering
\includegraphics[width=0.75\textwidth]{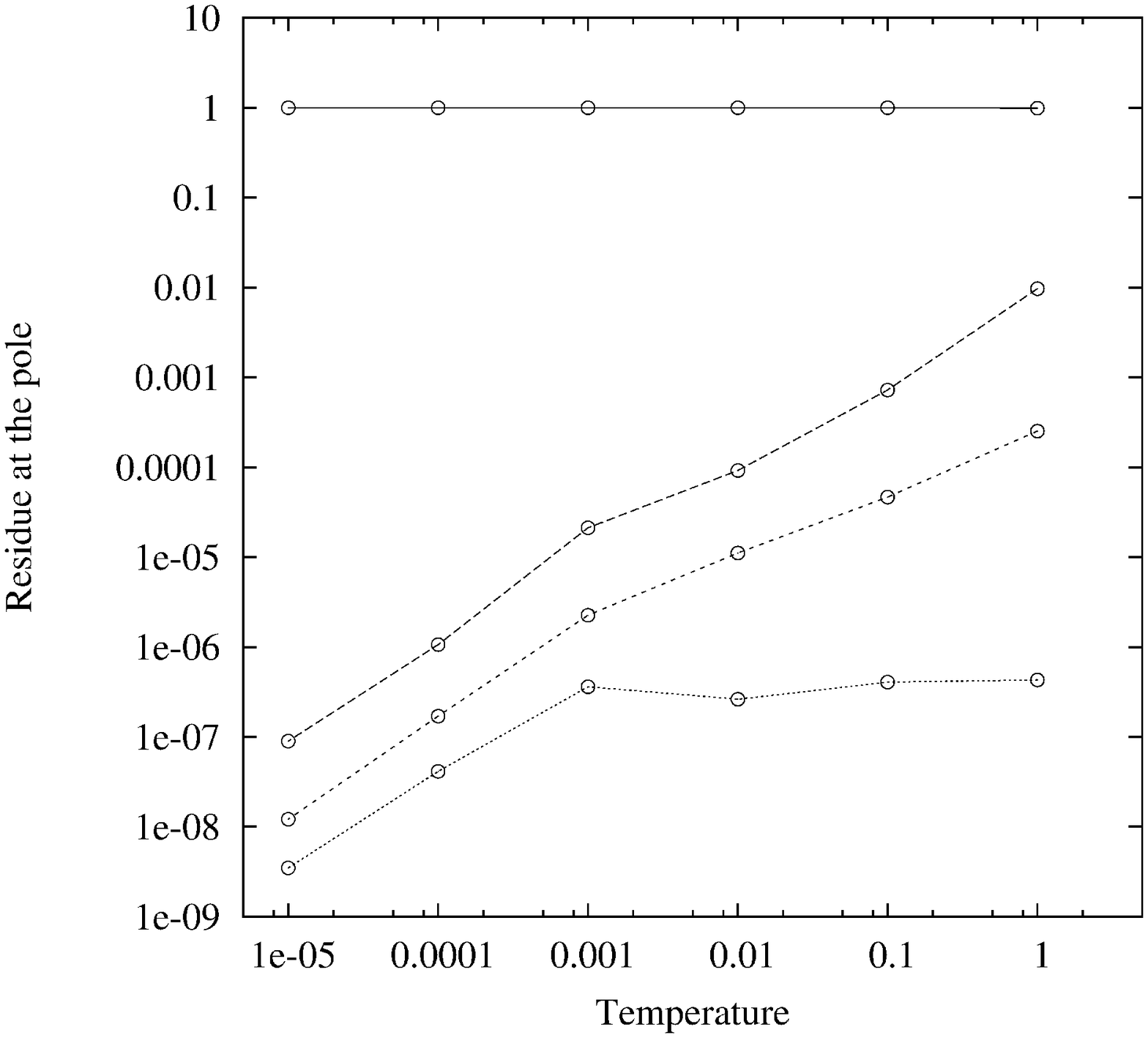}
 \caption{ \label{fig2} Residues at  the four poles, whose imaginary
    parts are reported in fig.~\ref{fig1} versus
    temperature, in logarithmic scale. Here the number of particles is $N=40$.}
\end{figure}

Let us describe the results in some detail.  We know that the rational
approximation of the Laplace transform at
order $2n$ will have $2n$ complex conjugated poles,
$s_k=i\omega_k$, $\bar{s}_k=-i\omega_k$ say.  The inverses of the
$\omega$'s so found for the time autocorrelation of $\tilde{\mathcal{E}}$ are
plotted in fig.~\ref{fig1} versus temperature $T$.   
The pole that shows up at order 1 produces the
upper curve in the figure.  It corresponds to oscillation of a quite
large period, growing as $1/T^{1/2}$ as $T$ decreases to zero.  The
actual value changes a little when the approximation increases, keeping
however the same order of magnitude.  The plotted data correspond to
the approximation order $n=4$. We have also noted that the values
found do not significantly change with the number $N$ of particles,
for $N$ up to 364. The frequencies represented by the 
lower curves show up at successive orders $n=2,3,4$, respectively, and
correspond to oscillations of a short period, which, at variance with
the first period, are of the order of magnitude of the typical
periods of the normal modes.  They appear to remain almost constant with $T$.

A further information is provided by the residues
of the poles, which represent the amplitudes of the
oscillations.  The quantities for the four poles of fig.~\ref{fig1} are
reported in fig.~\ref{fig2}, after a normalization defined by setting the sum of
the residues equal to one.  The remarkable fact is that the amplitude
corresponding to the longest period is the largest one.  Even for
$T=1$ the amplitudes of the shorter periods do not exceed one
percent of the total.

\begin{figure}[th]
 \centering
\subfloat[][Inverse of the imaginary part of the poles versus
  temperature, in logarithmic scale.\label{fig3}]{ \includegraphics[
    width=.45\columnwidth]{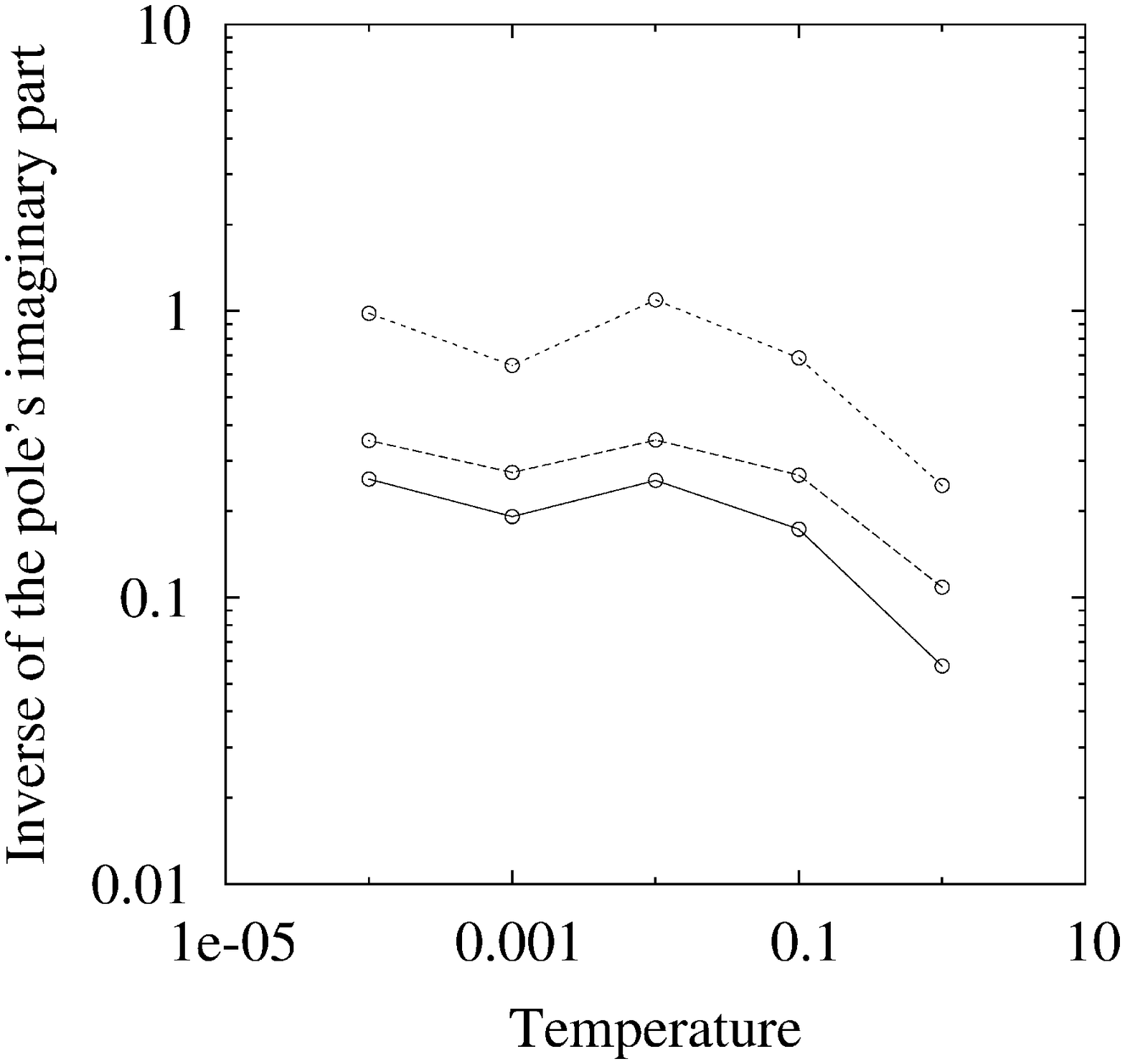}}\quad
\subfloat[][Residues at  the poles versus
    temperature, in logarithmic scale. Here the number of particles is
    $N=40$.\label{fig4}]{
  \includegraphics[width=.45\columnwidth]{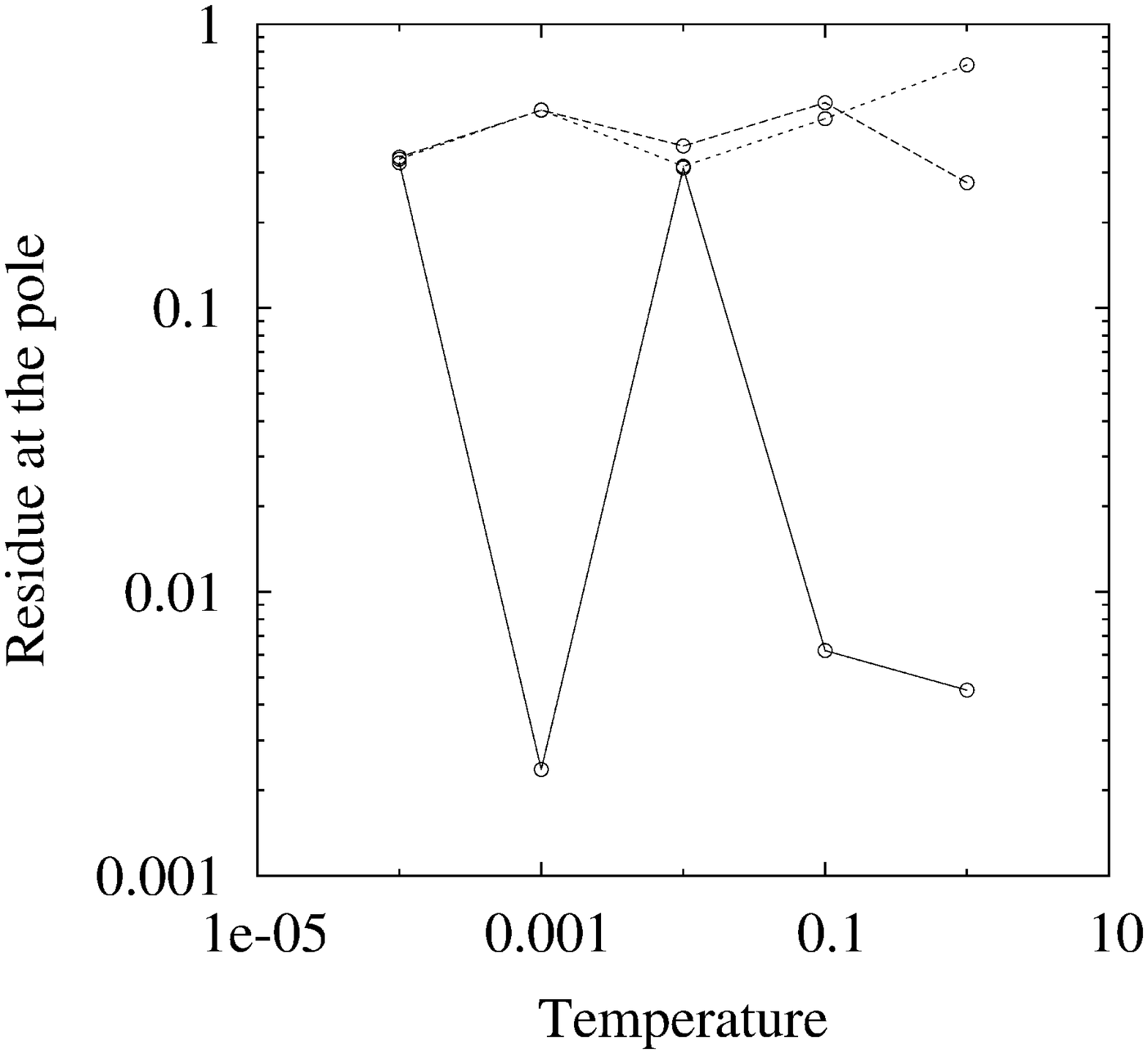}}
\caption{Location and residues of the first three poles of the Laplace transform
  of the time autocorrelation of kinetic energy $\tilde K$ versus
  temperature, in logarithmic scale. Here the number of particles is
    $N=40$.}
\label{fig34}
\end{figure}

It is now of interest to look at the corresponding graphs
(fig.~\ref{fig3}-\ref{fig4}) for the time
autocorrelation of the kinetic energy of half of the chain, $\tilde
K$. As expected, the time-scales involved are much shorter, as is
seen by inspection of fig.~\ref{fig3}. 
In any case, an indication of a bigger chaoticity is given by the fact that
all the residues are of the same order of magnitude and are closer to
each other, as  can be
seen in fig.~\ref{fig4}. This means that in the spectrum of the time
autocorrelation of $\tilde K$ at least three frequencies are excited,
at variance with the spectrum of $\tilde{\mathcal{E}}$, for which just
a frequency is actually excited.

\begin{figure}
  \centering
  \includegraphics[width=0.8\textwidth]{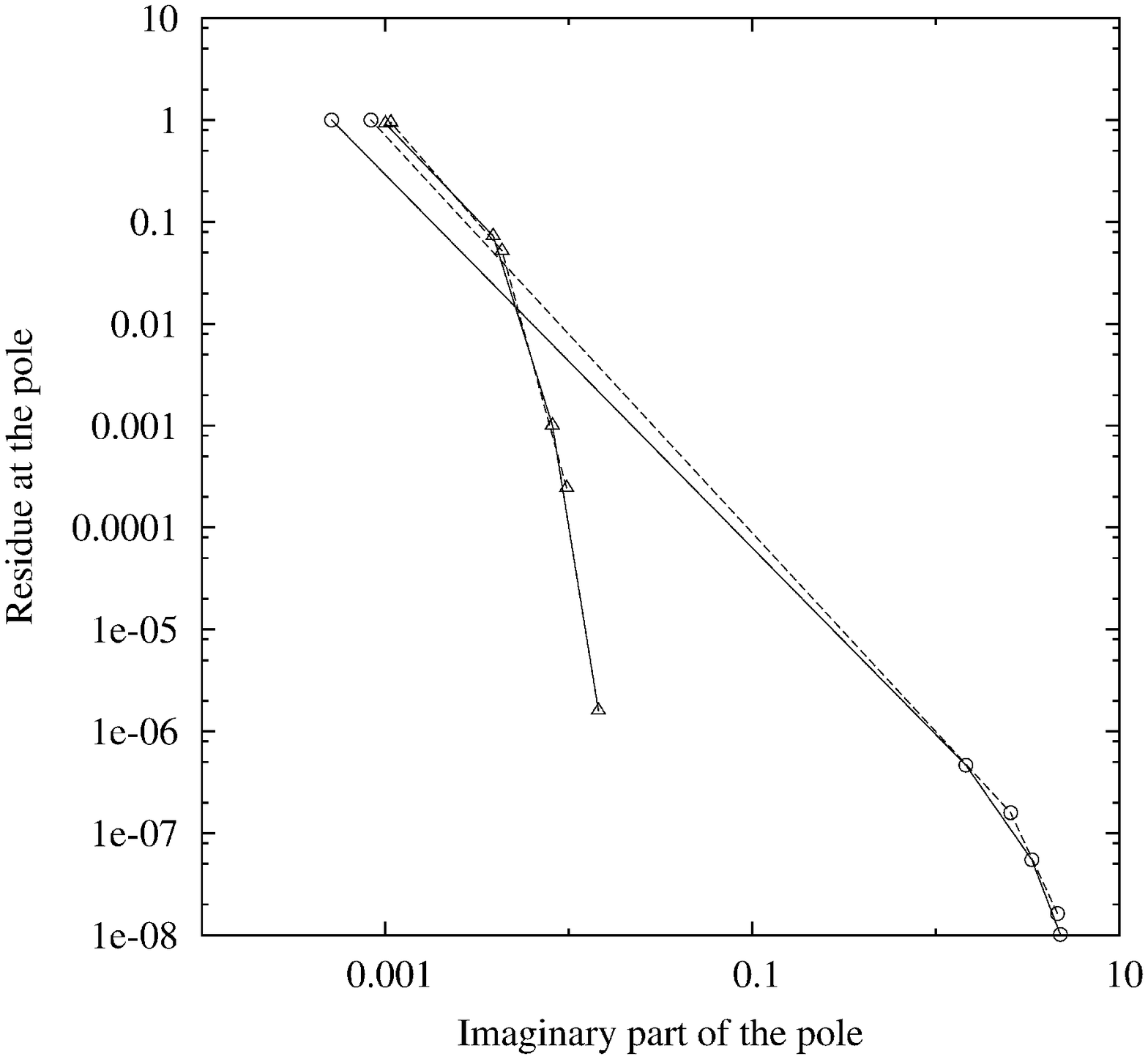}
  \caption{\label{fig5} Residues of the poles of the Laplace
    transforms of $\mathbf C_{\tilde{\mathcal{E}}}(t)$ and of $g(t)$ versus 
    their imaginary parts, in logarithmic scale. The circles refer to
    $\mathbf C_{\tilde{ \mathcal{E}}}(t)$ and the triangles to $g(t)$, while the
    dashed lines indicate that the order of approximation is 3, and
    the continuous lines indicate order 4. Here the number of particles is
    $N=40$ and $T=10^{-5}$}
\end{figure}

Then, we compare the behaviour of $\mathbf{C}_{\tilde{\mathcal{E}}}(t)$
with that of $g(t)\equal \cosh^{-1}(bt)$, where
$b>0$, in order to look for possible differences with respect to an exponential
decay. Here the constant $b$ is so chosen that the 
approximation of the Laplace transform of such a function at order $n=1$ has
a pole at the same place as the Laplace transform of $\mathbf C_{\tilde{
  \mathcal{E}}}(t)$, for a fixed temperature $T$ (as we have seen, $b\sim
1/T^{1/2}$). In fig.~\ref{fig5} the poles and the corresponding
residues of the Laplace transforms of  $\cosh^{-1}(bt)$ and $\mathbf C_{\tilde{
  \mathcal{E}}}(t)$ are plotted at the successive orders of approximation $n=3,4$.

In both cases a pole of order $10^{-3}$ has the largest residue, but it is
apparent that, while for $g(t)$ the other poles are close to that one, in the
case of $\mathbf C_{\tilde{\mathcal{E}}}(t)$ there seems to remain a
jump between the first pole and the others. This fact is in agreement with the
remark, already made by Stieltjes, that the poles of  the Laplace transform
of $g(t)$ become dense on the
imaginary axis  and corresponds with the fact that the spectral measure is in
this case analytic. On the other hand, the same fact seems to suggest
a qualitatively different behaviour of $\mathbf C_{\tilde{ \mathcal{E}}}(t)$, and
in particular that $\alpha(\omega)$ is not analytic in this case, so that the
decay of the time autocorrelation of $\tilde{\mathcal{E}}$ is not exponential.

As a final remark, we add a few considerations concerning the actual
difficulty occurring in the calculations.  In view of
Theorem~\ref{teor:stieltjes}, all 
poles lie on the imaginary axis and all the determinants
(\ref{eq:def_determinanti}) are positive. So, we are certain that the
outcome of our computation ceases to be reliable at the order at which
we encounter a negative determinant. Since the coefficients are determined
via a Monte Carlo approximation of an integral, they are subject to a
numerical error. The difficulty is that increasing
the order requires also a corresponding increase of the precision,
and so a longer and longer calculation.  This makes the whole procedure
unpractical even at not too high orders, and for this reason we
performed most calculations up to order 8 only, thus finding 4
frequencies.

\section{Conclusions}\label{sez:conclusione}
We have shown that there is a convergent algorithm which, in
principle, could be implemented to get the Laplace transform of the
time autocorrelation of a dynamical variable, and so to get
conclusions about the rate of its decay as $t\to \infty$. In our
opinion, the most relevant aspect is the connection we found
between the spectral measure and the solution of the Stieltjes moment
problem, which implies a relation between the smoothness property of
the measure and the fulfilment of an infinite set of constraints
(see, as an example, Proposition~\ref{prop:hausdorff} in
Appendix~\ref{app:hausdorff}). This leads us to the question, whether
smooth measures are in some sense exceptional, namely, to ask whether these
constraints can still be satisfied when small changes in the moments
are performed (i.e., by generic small perturbations of the Hamiltonian
flow).

Implementing such an algorithm on a computer raises major difficulties
with respect to both the computational time and the reliability of the
results. As a matter of fact, the main problem is connected with the
calculation of large determinants. However, we tried to apply the
method to two functions of interest fro the FPU problem, namely, the
energy of a packet of low frequency modes and the kinetic energy of
half of a FPU chain. For the first function, our results seem to
support the conjecture that the time autocorrelations exhibit a
sub-exponential decay, while no definite conclusions can be drawn for 
the second quantity.  We remark that this is a known characteristic
for chains of FPU type, namely that quantities related to the
particles typically exhibit a behaviour coherent with the predictions
of Statistical Mechanics, while this often does not happen with
quantities related to the modes. On the other hand, we should
emphasize that the existence of functions with a slow decay of the
autocorrelations is in itself an obstacle for the ergodic behaviour of
a dynamical system. For similar results, see, e.g., \cite{gpp}.

\appendix
\section{Conditions on the regularity of the measure which solves the
  moment problem}\label{app:hausdorff} 
The problem of determining a necessary and sufficient
condition in order that the solution of an Hamburger moment problem
admit a bounded positive derivative has been solved by KHz and
Krein (see \cite{ahikrein}). Unfortunately, their criterion
cannot be applied to study the derivative itself. Thus, we report the
criterion of Akhiezer and Krein for checking the existence of a
derivative for the Hamburger moment problem for the measure $\alpha$,
then we give an alternative procedure for studying the derivatives of
$\sqrt{u}\Phi'(u)$,\footnote{Recall that $\alpha'(\omega)$ is analytic
  if and only if $\sqrt{u}\Phi'(u)$ is (see
  Corollary~\ref{cor:analitico}).} at least in the case in which
the time autocorrelation is analytic on the whole real axis. We
convert the problem into an Hausdorff moment problem (i.e., a moment
problem on a finite interval), in order to get condition on the
regularity of the successive derivatives.

To state the result of \cite{ahikrein} of interest to us, let us
introduce the definition of a non-negative definite sequence of moments $c_n$,
by requiring that
$$
\sum_{i,j=1}^nc_{i+j}X_i\bar X_j\ge 0\ ,
$$
for any $n$ and any sequence $(X_1,\ldots,X_n)$ of complex
numbers.\footnote{Notice that this implies that the determinants of
  $\Delta_n$ defined in (\ref{eq:def_determinanti}) are all
  non-negative, but the converse is not true.} Then one has
\begin{proposizione}\label{prop:ahiezer}
In order that $\alpha'(\omega)$ exist and be bounded, it is necessary
and sufficient that there exist $L>0$ such that the sequence $t_0(L),
t_1(L),\ldots$ defined by\footnote{Recall that we have defined
  $c'_{2n}=c_n$ and $c'_{2n+1} = 0$.}
$$
t_k(L)=\frac1{(k+1)!L^{k+1}}\left|
\begin{array}{cccccc}
c'_0 & -L & 0 &\cdots & 0& 0\\
2 c'_1 & c'_0 & -2L &\cdots & 0 & 0\\
\vdots&\vdots &\vdots &\ddots&\vdots&\vdots\\
k c'_{k-1} &(k-1)c'_{k-2} &(k-2) c'_{k-3} &\ldots &c'_0&-kL\\
(k+1)c'_k & k  c'_{k-1} & (k-1) c_{k-2}&\ldots&2c'_1& c'_0\\
\end{array}
\right|\ ,
$$
is non-negative definite.
\end{proposizione}

For the study of the successive derivatives, we observe that in
paper \cite{hausdorff} Hausdorff found a solution method for the
moment problem on the interval  $[0,1]$, which also allows
to give a criterion to check whether, at any order, the
derivatives of the measure exist and are bounded. Such results can be
also applied to our problem if the time autocorrelation is analytic on
the whole real axis, because, through an analytic change of variable,
one can map $[0,1]$ on $\mathbb R^+$ and obtain the next
Proposition~\ref{prop:hausdorff}, which provides necessary and
sufficient conditions in order that the $n$-th derivative of
$\sqrt{u}\Phi'(u)$ exist and be bounded. For its statement, we need to
define the auxiliary moments $\mu_k$ and $\tilde\mu_k$:  we need to
preliminarily define
$$
\mathfrak{F}(z)\equal\int_0^{+\infty}\frac{\dif
  \Phi(u)}{z+u}\ ,
$$
then we set
\begin{equation}\label{eq:momenti_hausdorff}
\begin{split}
\mu_0\equal c_0\, \quad\mu_k\equal c_0-\frac
1{(k-1)!}\frac{\dif^{k-1}}{\dif^{k-1} z}\left(z^k 
\mathfrak{F}(z)\right)_{z=1}\ , \mbox{ for } k>0\ ,\\
\tilde\mu_k\equal \sum_{n=0}^{\infty}\binom{k+1/2}{n}\sum_{l= 0}^n
\binom{n}{l} (-)^{n-l}2^{l-k-1/2}\mu_l\ .
\end{split}
\end{equation}
These quantities are well defined, since it is known that  $\mathfrak{F}(z)$ is
analytic in $z=1$ and the binomial expansion converges.\footnote{See
  the discussion in the proof below.} Then, we
define the quantities
\begin{equation}\label{eq:definizioni_hausdorff}
\begin{split}
\lambda_{p,m}^0\equal \binom{p}{m}\sum_{k=0}^{p - m}
(-1)^k \binom{p-m}{k}\tilde{\mu}_k\ ,\quad\mbox{for }0\le m\le p\ ;\\
\lambda_{p,m}^k\equal (p+1)
\left(\lambda_{p,m}^{k-1}-\lambda_{p,m-1}^{k-1}\right)\ , \quad\mbox{for
}k>0, k\le m\le p-k\ ,
\end{split}
\end{equation}
and state the following
\begin{proposizione}\label{prop:hausdorff}
Let $\mathbf C_f(t)$ be an analytic function of $t$, for $t\in \mathbb
R$, and let $\Phi$ be the measure which solves the related Stieltjes
moment problem. A bounded derivative of order $n$ of
$\sqrt{u}\Phi'(u)$ exists if and only there exist $L_k>0$ such that,
for any $p>0$,  $0\le k<  
n$ and $k\le m\le p-k$, one has
\begin{equation}\label{eq:cond_hausdorff}
(p+1)\left|\lambda_{p,m}^k\right|\le L_k\ .
\end{equation}
\end{proposizione}
\textbf{Proof.}
The proof is performed through the Euler transformation
$w\equal \frac u{u+1}$. This enables one to define the positive
Borel measure $\Psi(w)$ on $[0,1]$, by $\Psi(w)\equal
\Phi(w/(1-w))$. The relation between the moments of $\Psi$
on the interval $[0,1]$ and the moments $c_n$ of $\Phi$ can be
obtained by the following chain of equalities, for $k>0$,
\begin{equation}\label{eq:hausdorff_1}
\int_0^1 w^k\dif \Psi(w)=\int_0^{+\infty}\frac{u^k}{(u+1)^k}\dif
\Phi(u)=\sum_{l=0}^k(-)^k \binom{k}{l} \int_0^{+\infty}\frac{\dif
  \Phi(u)}{ (u+1)^l}\ ,
\end{equation}
as can be seen by expanding $u^k=(u+1-u)^k$ via the binomial
formula. It is easy to see that this implies that the coefficients $\mu_k$
defined by (\ref{eq:momenti_hausdorff}) are the moments of $\Psi$.

In order to study the derivatives of $\alpha'(\omega)=2\sqrt u
\Phi'(u)$, we pass to the the moments given by the
distribution function $\sqrt w \Psi'(w)$. This function exists 
if $C_f(t)$ is analytic on the whole real
axis, because in such a case its spectral measure
$\alpha(\omega)$ tends exponentially fast towards its limit for
$\omega\to \infty$ and, in consequence, $\Psi(w)$ tends towards the
same limit as $w\to 1$ at least as fast as $e^{-\kappa/\sqrt{1-w}}$.
Moreover, $\sqrt w \Psi'(w)$ admits a bounded derivative of order $n$
if and only if $\sqrt u \Phi'(u)$ does. We need, thus, to find the
moments of $\sqrt w \Psi'(w)$, which are given by
$$
\int_0^1w^{k+1/2}\Phi'(w)\dif w= \int_0^1
\sum_{n=0}^{\infty}\binom{k+1/2}{n} 2^{k+1/2-n}\left(
w-\frac12\right)^n\Phi'(w)\dif w \ ,
$$
in view of the binomial expansion of $\sqrt{1/2+(w-1/2)}$. Since this
expansion converges uniformly for $w\in [0,1]$, one can exchange the
order of summation and integration and prove that these moments are
indeed the $\tilde \mu_k$ defined by (\ref{eq:momenti_hausdorff}).

As Hausdorff proved, in order that the moment problem for the sequence
$\tilde \mu_k$ have a solution $\psi_0$ on $[0,1]$, which is almost everywhere
differentiable with a limited derivative (in our case, it is $\psi_1(w)=\sqrt{w}
\Psi'(w)$), it is necessary and 
sufficient that $(p+1)|\lambda^0_{p,m}| $ be bounded, where
$\lambda^0_{p,m}$ are defined by equation
(\ref{eq:definizioni_hausdorff}). This corresponds to
$$
\lambda^0_{p,m}= \binom{p}{m}\int_0^1w^m(1-w)^{p-m}\dif \psi_0(w)\ .
$$
Then, one considers the Hausdorff moment problem in which the measure
(not necessarily positive) is $\psi_1$, and expresses its moments
$\tilde\mu^1_k$ through the original moments $\tilde \mu_k$, by integrating by
parts. So one obtains that the coefficients $\lambda_{p,m}^1$ defined by
(\ref{eq:definizioni_hausdorff}) play for $\psi_1$ the same role which
the $\lambda_{p,m}^0$ play for $\psi_0$, because, for $1\le m\le
p-1$,\footnote{We neglected here the terms for $m=0$ and $m=p$, since
  they give indications only on the discontinuities of $\psi_1$, which
  can be disposed of, since $\psi_1$ can be defined at will on a set of
  measure zero.}
$$
\lambda^1_{p,m}= (p+1)(\lambda^0_{p,m}-\lambda^0_{p,m-1})=
\binom{p+1}{m}\int_0^1w^m(1-w)^{p+1-m}\dif \psi_0(w) \ .
$$
Thus, one can express the condition that $\psi_1$ admits a limited
derivative by asking that $(p+1)|\lambda^1_{p,m}|$ have an upper
bound. The proof is then concluded by iterating this way of proceeding.
\begin{flushright}Q.E.D.\end{flushright}
\blank

Finally, we add the following Proposition, which can be proved by
the easy remark that a power series with positive coefficients, which
has a radius of convergence $\rho$, has a singularity for $z=\rho$
(this is known as Vivanti's theorem). 
\begin{proposizione}\label{prop:convergente}
In order that the decay of the time autocorrelation of the dynamical
variable $f$ be not 
exponentially fast it is sufficient that
there exists $\rho>0$ such that
  $\limsup_{n\to+\infty}\sqrt[n]{c_n}\le\rho$\ .
\end{proposizione}

\noindent
\textbf{Remark 5.} The condition of Proposition~\ref{prop:convergente}
  requires only to have an upper bound 
on the $c_n$, whereas those of
Propositions~\ref{prop:ahiezer}-\ref{prop:hausdorff}  ask for a more detailed
knowledge on the coefficients $c_n$. We believe that the first one is
seldom fulfilled (an example 
  in which it is fulfilled is that of a harmonic oscillator).  For,
  the $n$-th derivative of a function usually grows as $n!$, so that 
$$
\limsup_{n\to+\infty}\sqrt[n]{c_n}=+\infty\ .
$$
The other requirements enables us to deal with a larger class of
variables, but, as just said, it has the drawback of needing a very detailed
knowledge of all coefficients.

\section{Proof of Proposition~\ref{prop:a_priori}}\label{app:a_priori} 
Consider the series (\ref{eq:serie_troncata}), and
recall that the remainder has a definite sign.  Due to the inequalities
(\ref{eq:correlazione_e_varianza}) we readily get the bounds
$$
a^0_0+\sum_{k=1}^{2n} (-)^k a^0_k t^{2k}\ge -a^0_0 \quad \mbox{and}\quad
a^0_0+\sum_{k=1}^{2n+1} (-)^k a^0_kt^{2k}\le a^0_0\ ,\quad \forall t\in\mathbb{R}\ ,
$$ 
which hold true because the remainder is negative in the first case
and positive in the second one.  Setting $y\equal t^2$, we get the
inequalities
\begin{eqnarray}
P_n(y) &\equal & 2 a^0_0+\sum_{k=1}^{2n}(-)^ka^0_k y^k\ge 0 \ , 
\label{eq:richiesta_coefficienti_P}\\
Q_n(y) &\equal  &\sum_{k=0}^{2n}(-)^ka^0_{k+1} y^k\ge 0 \ , 
 \label{eq:richiesta_coefficienti_Q}
\end{eqnarray}
i.e., the polynomials $P_n(y)$ and $Q_n(y)$ are
positive semi-definite for $y\ge 0$.  Look now for the real roots of
the equations $P_n(y)=-\eps$ and $Q_n(y)=-\eps$, with
$0<\eps<\min\{a_0,a_1\}$.  Since $P_n(y)$ and $Q_n(y)$ are polynomials
of degree $2n$ in which $2n$ changes of sign occur, Descartes' rule of
signs implies that such roots, if they exist, must be positive.  Thus
$P_n$ and $Q_n$ are positive semi-definite on the whole real axis,
i.e., the inequalities (\ref{eq:richiesta_coefficienti_P}) hold true
for all $y\in \mathbb{R}$.

\noindent
Consider now the dynamical variable $f^{(l)}$, calculated by recursive
application of the Poisson bracket with $H$.  By a straightforward
application of Theorem~\ref{teor:serie} we get that the coefficients of
$f^{l}$ are precisely  $a_n^l$ as defined
in~(\ref{eq:coefficienti_generalizzati}).  Since the argument above
applies to any dynamical variable, then we get that the corresponding
polynomials  $P_n^l(y)$ and
$Q_n^l(y)$ are positive semi-definite, too, as claimed.

\section{Computation of the coefficients needed for the Fa\`a di Bruno
  formula}\label{app:faa} 

Let us recall the Fa\`a di Bruno's formula (see \cite{faadibruno}).
Let a function  $f\bigl(g(x)\bigr)$ be given, where $f$ and
$g$ possess a sufficient number of derivatives.  Then one has
$$
\frac{\dif^n}{\dif x^n} f\bigl(g(x)\bigr)
  =\sum \frac{n!}{k_1!\,k_2!\,\cdots\,k_n!}\cdot
  f^{(k_1+\cdots+k_n)}\bigl(g(x)\bigr)\cdot 
   \prod_{j=1}^n\left(\frac{g^{(j)}(x)}{j!}\right)^{k_j}\  ,
$$
where the sum is over all $n$-tuples of nonnegative integers
$(k_1,\ldots,k_n)$ satisfying the constraint
$$
1\cdot k_1+2\cdot k_2+3\cdot k_3+\cdots+n\cdot k_n=n \ .
$$

In our case, we apply twice the formula, the first time to find the
successive derivatives of the canonical coordinates $q$ and $p=\dot q$
with respect to time, the second one to find the derivatives of
$\tilde{\mathcal E}(p,q)$ and $\tilde K(p,q)$. Whereas the latter
application is trivial, we remark that the former is
obtained by explicitly expressing the force $\ddot q=\dot p$ as a
function of $q$ and $p$, then by writing
$$
\frac{\dif^n}{\dif t^n}q=\frac{\dif^{n-2}}{\dif t^{n-2}}\ddot q(p,q)\ .
$$
As the term at the r.h.s. involves only derivatives of $q$ up to
order $n-1$, the scheme can be applied iteratively to obtain all the
derivatives of $q$ (and $p$) with respect to time.

Since the derivatives of $f$ and $g$ are easily calculated in our
case, the problem is just to have an effective algorithm that produces
all the required $n$-tuples $(k_1,\ldots,k_n)$.

For positive integers $n,s$ we introduce the sets
$$
\Kscr_{n,s} = \{k\in\naturali^n_{0}\>:\>k_1+2k_2+\ldots+nk_n=s\}
$$
where $\naturali_{0}$ is the set of non negative integers.  For $n=1$
we have $\Kscr_{1,s} = \{(s)\}$, i.e., just one element.  For $n\gt 1$
the following statement is obviously true: {\sl every element
$(k_1,\ldots,k_n)\in\Kscr_{n,s}$ can be written as $(k', k_n)$ where
$k'=(k_1,\ldots,k_{n-1})\in\Kscr_{n-1,s-nk_n}\,$.}  This is the key of
our algorithm.

In $\Kscr_{n,s}$ we consider the inverse lexicographic order, right to
left.  More precisely, the ordering is recursively defined as follows:
for $n=1$ the ordering is trivial, since there is just one element;
for $j,k\in\Kscr_{n,s}$ with $n\gt 1$ we say that $j\lt k$ if either
case applies: (i)~$j_n\gt k_n\,$, or (ii)~$j_n=k_n$ and
$(j_1,\ldots,j_{n-1})\lt (k_1,\ldots,k_{n-1})$.  E.g., setting $n=3$
and $s=5$ we get the ordered set
$$
\bigl\{(0, 1, 1)\,,\>
(2, 0, 1)\,,\>
(1, 2, 0)\,,\>
(3, 1, 0)\,,\>
(5, 0, 0)
\bigr\}\ .
$$
Remark that the last vector $k\in\Kscr_{n,s}$ is $(s,0,\ldots,0)$.

We come now to stating the algorithm.  We use two basic operations,
namely: (a)~find the first vector $k\in\Kscr_{n,s}$ according to the
ordering above; (b)~for a given $k\in\Kscr_{n,s}$ find the next
vector.

The first vector is easily found by setting
$$
k_n = \left\lfloor\frac{s}{n}\right\rfloor\>,
 \ k_{n-1} =  \left\lfloor\frac{s-nk_n}{n-1}\right\rfloor\>,
  \ldots,
   \ k_{1} =  s-nk_n-\ldots-2k_2\ .
$$
For a given $k$ the next vector is found as follows: starting from
$k_2$ find the first $k_m$ which is not zero, so that
$k_2=\ldots=k_{m-1}=0$ and $k_m\ne 0$.  Then replace $k_m$ with
$k_m-1$ and $(k_1,\ldots,k_{m-1})$ with the first vector of
$\Kscr_{m-1,k_1+m}\,$.  The algorithm stops when $k=(s,0,\ldots,0)$,
namely the last vector, because the index $m$ can not be found.


\begin{thebibliography}{}
\bibitem{liverani} C. Liverani, Ann.
  Math. \textbf{159}, (2004) 1275--1312.
\bibitem{chernov} N. Chernov, R. Markarian, \textit{Chaotic
  billiards}, AMS (Providence, 2006).
\bibitem{carati} A. Carati, J. Stat. Phys. \textbf{128}, (2007) 1057--1077.
\bibitem{maiocchicarati} A.M. Maiocchi, A. Carati, Commun. Math. Phys.
  \textbf{297}, (2010) 427--445.
\bibitem{caratimaiocchi} A. Carati, A.M. Maiocchi, 
  Commun. Math. Phys. \textbf{314}, (2012) 129--161.
\bibitem{koopman} B.O. Koopman,  Proc. Natl. Acad.
   Sci. USA \textbf{17}, (1931) 315--318.
\bibitem{obre} N. Obre\v{s}kov, \textit{Verteilung und Berechnung der
  Nullstellen reeller Polynome}, VEB Deutscher Verlag der
  Wissenschaften (Berlin, 1963).
\bibitem{ruelle} D. Ruelle,
  Phys. Rev. Lett. \textbf{56}, (1986) 405--407.
\bibitem{stieltjes} T. J. Stieltjes, Ann. Fac. Sci. Toulouse S\'er. 1
  \textbf{8}, (1894) J1--J122  and \textbf{9},
  (1895) A5--A47. 
\bibitem{widder} D.V. Widder, \textit{The Laplace transform}, Princeton
  University Press  (Princeton, 1946).
\bibitem{feller} W. Feller, \textit{An introduction to probability theory and
  its applications}, vol. II., Wiley \& Sons (New York, 1971).
\bibitem{ahiezer} N.I. Akhiezer, \textit{The classical moment problem and
  some related questions in analysis}, Oliver \& Boyd  (Edinburgh, 1965).
\bibitem{hausdorff} F. Hausdorff, Math. Z. \textbf{16}, (1923) 220--248.
\bibitem{fpu} E. Fermi, J. Pasta, S. Ulam, in \textit{E. Fermi Collected 
Papers}, vol. 2, pp. 977--988, The University Chicago Press (Chicago, 1965).
\bibitem{cggp} A. Carati, L. Galgani, A. Giorgilli, S. Paleari,
  Phys. Rev. E \textbf{76}, (2007)  022104. 
\bibitem{gpp} A. Giorgilli, S. Paleari, T. Penati, Extensive adiabatic
  invariants for nonlinear chains, preprint.
\bibitem{ahikrein} N.I. Akhiezer, M.G. Krein, \textit{Some questions in the
  theory of moment}, AMS  (Providence, 1962).
\bibitem{faadibruno} F. Fa\`a di Bruno,
  Ann. Sci. Mat. Fis. \textbf{6}, (1855) 479--480. French version
in: Note sur une nouvelle formule de calcul diff\'erentiel, Quart. J. Pure Appl.
Math. \textbf{1}, (1857) 359--360.
\end{thebibliography}
\end{document}